\documentstyle[preprint,aps]{revtex}

\newcommand{\be}{\begin{equation}}
\newcommand{\ee}{\end{equation}}
\newcommand{\bea}{\begin{eqnarray}}
\newcommand{\eea}{\end{eqnarray}}

\tightenlines

\begin{document}

\draft
%  \twocolumn[\hsize\textwidth\columnwidth\hsize\csname
%  @twocolumnfalse\endcsname

\title{\bf Studies of the $t\hbox{-}J$ two-leg ladder via series expansions
}
\author{J. Oitmaa\cite{byline1}, C.J. Hamer\cite{byline2}, and Zheng Weihong\cite{byline3}}
\address{School of Physics,
The University of New South Wales,
Sydney, NSW 2052, Australia.}

%\date{July 26, 1999}
\date{\today}

\maketitle

\begin{abstract}
Series expansions at $T=0$ are used to study properties of the half-filled $t\mbox{-}J$
ladder doped with one or two holes, and at quarter filling.
Dispersion curves are obtained for one-hole symmetric and
antisymmetric (bonding and antibonding) excitations and for
the two-hole bound state.  The line in the phase diagram that separates bound and
unbound states is determined. For quarter filling we compute
the ground state energy and estimate the location of the phase separation line.
Comparisons with other numerical and analytical results are
presented.
\end{abstract}

% \phantom{.}
% ]

\narrowtext
\section{INTRODUCTION}
The discovery of materials containing $S={1\over 2}$ ions
which form a 2-leg ladder structure\cite{azu94,hir95} has
stimulated a great deal of theoretical work to understand the unique
and surprising features of these systems (for reviews see \onlinecite{dag96,ric97}).
Early work on magnetic properties confirmed the existence of a spin-gap
in the 2-leg ladder with antiferromagnetic (AF) interactions, and
hence a qualitative difference between even and odd legged spin ladders.
More exotic spin ladders, with frustration, dimerization and
other features have been studied more recently.\cite{mar96,ladder,ner98,cab,kot99}.

Even more interesting behaviour may be expected if the system includes charge
degrees of freedom. This can be achieved by doping, to  create a system
of strongly correlated mobile holes, as in the cuprate superconductors.
Indeed an understanding of this (arguably simpler) system may yield important consequences
for a better understanding of high $T_c$ superconductivity. The simplest
model, in this context, is the $t\mbox{-}J$ model away from half-filling.
Considerable work has been done on the $t\mbox{-}J$ model on a 2-leg ladder, through
exact diagonalizations\cite{dag92,tsu94,poi95,hay96,hay962,mul98,rie99}, the
density matrix renormalization group approach\cite{whi97,sie98},
and approximate analytic theories\cite{sie98,sig94,lee,sus98}.
A number of broad features have emerged from this work.

The $t\mbox{-}J$ ladder is shown in Figure 1. The Hamiltonian is
\bea
H = && J \sum_{i,a} ( {\bf S}_{i,a} \cdot {\bf S}_{i+1,a}
   -{1\over 4} n_{i,a} n_{i+1,a} ) 
  + J_{\perp} \sum_{i} ( {\bf S}_{i,1} \cdot {\bf S}_{i,2}
   -{1\over 4} n_{i,1} n_{i,2} ) \nonumber \\
 && - t \sum_{i,a,\sigma} P (c^{\dag}_{i,a,\sigma} c_{i+1,a,\sigma} 
        + H.c. ) P  - t_{\perp} \sum_{i,\sigma} P (c^{\dag}_{i,1,\sigma} c_{i,2,\sigma} 
        + H.c. ) P  \label{eqH}
\eea
where $i$ runs over rungs, $\sigma$ ($=\uparrow$ or
$\downarrow$) and $a$ (=1,2) are spin and leg indices,
 $c^{\dag}_{i,a,\sigma}$ ($c_{i,a,\sigma}$) is the creation
(annihilation) operator of an electron with spin $\sigma$ 
 at the $i$th site of the $a$th chain, and 
${\bf S}_{i,a}= {1\over 2} c^{\dag}_{i,a,\alpha} 
\sigma_{\alpha,\beta} c_{i,a,\beta} $
 denotes the $S=\case 1/2$ spin operator at the $i$th site of the $a$th chain.
 $P$ is a projection
operator that ensures that doubly occupied states are
excluded. %Unless noted otherwise we set $t=t_{\perp}$.

To understand the nature of the ground state and excitations of the system it
is helpful to consider the limiting case $J_{\perp}, t_{\perp} \gg J,t$
when the states on the rungs of the ladder form an appropriate basis.
Many of these features persist even to the isotropic case $J_{\perp}=J$, $t_{\perp}=t$.
At half-filling the charge degrees of freedom are frozen out and the system
is equivalent to the Heisenberg spin ladder. The ground state is a gapped spin
liquid with each rung in a spin singlet state. Spin excitations originate from exciting
one or more of the rungs to triplet states.

When the system is doped, leading to removal of electrons, all nine states of 
a rung must be included. These are discussed in Ref. \onlinecite{tsu94}, and for
completeness are illustrated in Table I. A single hole (of either spin) will go
into a bonding or antibonding state on a rung and this state can propagate along the
ladder. Troyer {\it et al.}\cite{tsu94} have obtained one-hole excitation spectra
from exact diagonalizations of systems up to 20 sites. They argue that the free
spin created remains bound to the hole and the resultant quasiparticle carries
both spin and charge. This is unlike the situation in the $t\mbox{-}J$ chain where
spin-charge separation occurs, in accordance with Luttinger liquid
theory.

Many authors have considered two-hole states. It is energetically favourable
for the two holes to form a bound state, on the same rung. The lowest two-hole
excitation  arises from a coherent propagation of the two holes along the
ladder. This excitation carries charge $2e$ and no spin. There are other,
more complex, excitations which are described by Troyer {\it et al.}\cite{tsu94}.

The nature of the system at finite doping is also of considerable interest. Poilblanc
{\it et al.}\cite{poi95} have conjectured a $T=0$ phase diagram in the plane of $J/t$
vs electron density $n$. For large $J$ one expects phase separation,
where the system separates into hole-free and hole-rich regions.
For smaller $J/t$ and at or near half-filling, there is a ``C1S0"
phase\cite{bal96}, with
one gapless charge mode and no gapless spin modes.
The spin-gap will decrease, as $n$ is decreased from 1, and a transition 
to a Luttinger liquid phase is predicted. At small $n$ electron-pairing 
rather than hole-pairing will occur. M\"uller and Rice\cite{mul98} have
extended this phase diagram through the inclusion of a
Nagaoka phase at low doping and small $J$, and a ``C2S2" phase with gapless charge
and spin excitations at larger $J$, between the Nagaoka and
spin-gapped phases. Much of this remains somewhat speculative.

In this paper we study the $t\mbox{-}J$ ladder via the method of series expansions.
This approach is complementary to other numerical methods and is able
to provide ground state properties, quantum critical points,
and excitation spectra to high accuracy. Its other advantage is that one 
deals, at once, with a large system and hence there are no finite size 
corrections needed. We have used this approach recently\cite{ham98} in a study of
the $t\mbox{-}J$ model on the square lattice, and we refer to that paper for
details of the method and references to previous works. 
Early perturbative studies of the $t\mbox{-}J$ model\cite{bar92} only considered
the lowest order terms.
The emphasis of our work is on 
one- and two-hole states, and we display excitation spectra for these. The
method is not well suited to handle finite doping and hence we are
not able to add to knowledge of the phase diagram for general $n$. We do however consider
the special case of quarter-filling, and obtain an estimate of the 
boundary of phase-separation in this case. Wherever possible we compare our results
with previous work.

This paper is organized as follows. In Sec. II, we consider the system with 
one and two holes. In Sec. III, we study the system at quarter-filling.
The last section is devoted to discussion and conclusions.

\section{One-hole and two-hole states}

The series expansion method is based on a linked cluster formulation of
standard Rayleigh-Schr\"odinger perturbation theory, with the
Hamiltonian (\ref{eqH}) written in the usual manner as
\be
H = H_0 + x V
\ee
where $H_0$ is a solved, unperturbed system and  $xV$ represents the  remaining terms.
The most obvious choice is to include all of the rung terms in $H_0$
\bea
H_0 &=&  J_{\perp} \sum_{i} ( {\bf S}_{i,1} \cdot {\bf S}_{i,2}
   -{1\over 4} n_{i,1} n_{i,2} ) 
   - t_{\perp} \sum_{i,\sigma} P (c^{\dag}_{i,1,\sigma} c_{i,2,\sigma} 
        + H.c. ) P 
\eea
and the coupling between rungs as $V$
\bea
x V &=& J \sum_{i,a} ( {\bf S}_{i,a} \cdot {\bf S}_{i+1,a}
      -{1\over 4} n_{i,a} n_{i+1,a} )  - t \sum_{i,a,\sigma} P (c^{\dag}_{i,a,\sigma} c_{i+1,a,\sigma} 
        + H.c. ) P  
\eea

As discussed above, the eigenstates of $H_0$ are direct products constructed
from the nine possible rung states, shown in Table I. To compute
the perturbation series we fix the values of $t_{\perp}/J_{\perp}$ and
$t/J$ and derive an expansion in powers of $x\equiv J/J_{\perp}$ for
the quantity of interest. The series is then evaluated (summed) at
the desired value of $J/J_{\perp}$ by using standard Pad\'e approximants and
integrated differential approximants\cite{gut}. We set $J_{\perp}=1$ to 
define the energy scale. To improve the convergence of the expansion we also
try the effect of adding a term to $H_0$ and subtracting it from $V$, for example
\bea
H'_0 &=&  (J_{\perp} + r) \sum_{i} ( {\bf S}_{i,1} \cdot {\bf S}_{i,2}
   -{1\over 4} n_{i,1} n_{i,2} ) 
  - t_{\perp} \sum_{i,\sigma} P (c^{\dag}_{i,1,\sigma} c_{i,2,\sigma} 
        + H.c. ) P \\
x V' &=& x V - r \sum_{i} ( {\bf S}_{i,1} \cdot {\bf S}_{i,2}
   -{1\over 4} n_{i,1} n_{i,2} ) 
\eea
and adjusting $r$ to obtain best convergence. Other types of term
could be added and subtracted in this way.

A slightly different formulation takes only the rung spin Hamiltonian
as the unperturbed part
\be
H_0 = J_{\perp} \sum_{i} ( {\bf S}_{i,1} \cdot {\bf S}_{i,2}
   -{1\over 4} n_{i,1} n_{i,2} ) \label{h07}
\ee
with all of the remaining terms, including the rung hopping,
included in the perturbation. Both methods have been used, as appropriate.

The series method can be used to compute the ground state energy and other
ground state properties such as magnetization, susceptibility, and correlations.
In the present study we are mainly interested in the nature of excitations.
Rather than calculate the ground state energy we have used the method
of Gelfand\cite{gelfand} to compute the excitation energy directly. For the 
one-hole excitations we start with an initial state with the hole in
either the bonding or antibonding state  on a particular rung and 
all other rungs in spin singlet states. The perturbation will then mix
this with other states in which the hole moves along  the ladder within
a fluctuating spin background. For the two-hole case the initial state has 
both holes on the same rung. During the evolution of the system these
holes may separate. Those final states which have the holes together
on the same final rung contribute to the two-hole excitation spectrum.
The series have been computed to order $(J/J_{\perp})^{12}$ for
the one hole case and to order $(J/J_{\perp})^{15}$ for the two-hole case.
In special cases we have obtained considerably longer series.
In the case of static holes $(t=0)$ we have extended the series to order
$(J/J_{\perp})^{21}$ for both one- and two-hole cases, while
for $J=0$ we have obtained one-hole series to $(t/J_{\perp})^{19}$
and two-hole series to $(t/J_{\perp})^{22}$.

The series are available on request. For the interest of the reader we give
some of the low order terms for both one and two-hole series in Appendix A.
As usual with series methods the convergence is judged by consistency
of different Pad\'e (or integrated differential)  approximants. In the present case convergence worsens for
increasing $J/J_{\perp}$ and $t/J_{\perp}$. Error estimates are shown in the figures 
displaying our results.

Figure 2 shows the excitation spectra for one-hole symmetric and
antisymmetric states for $J/J_{\perp}=0,0.5,1.0$ and various $t/J_{\perp}$.
To avoid excessive data we have set $t=t_{\perp}$
throughout. The quantities graphed are the gaps $\Delta E^{1S}/J_{\perp}$,
$\Delta E^{1A}/J_{\perp}$, where for instance
\be
\Delta E^{1S} = E^{1S} - E_0
\ee
and $E_0$ is the energy of the half-filled ground state. 
We draw attention to the following features.

\begin{itemize}
\item The series yield smooth dispersion curves throughout  the 1-d Brillouin zone.
There is consistency between different approximants and we believe
the error is small. The only exception is for some of the antisymmetric 
states where the figure shows larger error bars.

\item For fixed $J/J_{\perp}$ the energies decrease with increasing $t$.
The kinetic energy gain, proportional to $t$, exceeds the loss in potential energy from
breaking antiferromagnetic bonds.
% Increasing $t$ will increase the kinetic energy of a bare hole but the renormalization effect
% if the spin background more than conpensates for this, as can be seen from the 
% analytic theory.\cite{sus98}
The bandwidth also increases with increasing $t$, as expected.
The gap vanishes or goes negative when $t/J_{\perp}$ gets large,
i.e. the 1-hole state lies lower in energy than the half-filled ground state.

\item The main effect of increasing $J$ is in the position of the minimum.
For the symmetric states this is at $k=\pi$ for small $J$, and moves
to a value near $k=2 \pi/3$ for $J$ greater than a critical value
around $J/J_{\perp}\simeq 0.6$, almost independent of $t$.

\item The antisymmetric state shows greater sensitivity to the value of $J$, with a 
qualitative change in the one-hole dispersion curve between $J=0$ and 0.5.

\end{itemize}

These dispersion curves are in broad qualitative agreement with those obtained by
other methods\cite{tsu94,sus98}, but with significant quantitative differences. In Figure 3
we reproduce our results for the one-hole symmetric state spectrum for
$J/J_{\perp}=1$, and compare with the diagonalizations of Troyer {\it et al.}\cite{tsu94},
and an approximate analytic result of Sushkov\cite{sus98}.
The dispersion curve obtained from exact diagonalization has
a very similar shape to our one, but lies at somewhat lower energies.
The difference
may be due to finite size effects in the diagonalization results.
The analytic calculation, on the other hand, lies at high energies
and appears to overestimate the dispersion. The discrepancy is greatest
at $k=0$.  Note that ref. \onlinecite{sus98}
does not include the $- n_i n_j/4$ terms in the $t\mbox{-}J$ Hamiltonian and hence a 
constant 0.75 must be added.

We now turn to the two-hole excitations. Figure 4 shows the
excitation spectra, again for $J/J_{\perp}=0,0.5,1.0$ and for various values
of $t/J_{\perp}$. The quantity graphed is again the 2-hole gap
\be
\Delta E^{2h} = E^{2h} - E_0
\ee
where $E^{2h}$ is the energy of the lowest-lying singlet 2-hole state.
The following features are apparent

\begin{itemize}
\item The minimum now occurs at $k=0$ for all $t,J$.

\item The curves are quite flat for small $t/J_{\perp}$, as expected, and develop greater
dispersion and more structure as $t$ increases. The matrix element for two holes
to hop from one rung to the next is of order $t^2$.

\item The energy at $k=\pi$ depends only weakly on $t$. For the special case $J=0$
the energy at $k=\pi$ is in fact completely independent of $t$.
It is possible to prove this results quite simply by looking at matrix elements
of two-hole hopping processes. We do not have a simple physical explanation.

\item The two-hole gap vanishes or goes negative for small $k_x$ when $t/J_{\perp}$ gets 
large.

%\item We do not see any evidence for a linear dispersion relation near $k=0$,
%as discussed in Ref. \onlinecite{tsu94}.
\end{itemize}

The two-hole binding energy can be computed from our excitation energies by 
\be
E_b = \Delta E^{2h} - 2 \Delta E^{1h} = E^{2h} - 2 E^{1h} + E_0 \label{eq_binding}
\ee
where $\Delta E^{1h}$ is the lowest 1-hole energy, which is the symmetric one in 
our case. Negative $E_b$ corresponds to binding of the two holes.
Figure 5 shows $E_b$  versus $t/J_{\perp}$ for various values of $J/J_{\perp}$.
For $t=0$ all the curves converge to the vicinity of $-1$, which is the exact
(trivial) binding energy for $J=0$.
For larger $J$ this value increases slightly (to approximately -0.87 for large $J$),
but this  is below the resolution of this figure. For small $J$ the binding energy
decreases in magnitude for increasing $t$, and the two holes
no longer bind beyond a critical $t_c$, which is itself an increasing function of $J$.
For larger $J$ binding seems to occur for any $t$.
Our results are numerically consistent with those of Hayward {\it et al.}\cite{hay962}
and Troyer {\it et al.}\cite{tsu94}. For the isotropic case $J=J_{\perp}$,
the two hole state is bound for all values of $t$.

\section{The quarter-filled case}
We have obtained a number of results for the case of quarter filling, i.e.,
where there is one electron (or hole) per rung. This value of electron density
lies within the Luttinger liquid (C1S1) phase in the conjectured phase 
diagram\cite{poi95}, but rather near and parallel to the line separating
this phase from the spin-gapped phase. There has been recent interest in this
case as a model for NaV$_2$O$_5$\cite{smo98,nis98,nis99}.

To develop the perturbation expansions for quarter-filling we start from an
unperturbed state which is a direct product of alternating spin-up
and spin-down rung bonding states. 
% but still we have $2^{N/2}$ degenerate states since
% we have degenerate spin-up and spin-down bonding states in each rung
%(where $N$ is no. of sites). To remove this degeneration, 
In the absence of $V$ this is but one of $2^{N/2}$ degenerate states, because the
spin-up and spin-down rung bonding states have the same energy.
To remove this degeneracy, 
we have considered the effect of the perturbation $V$
on a system of two rungs, and have included the diagonal terms in the unperturbed
Hamiltonian.  This splits the degeneracy and favours a N\'eel type ordering
along the ladder.
To improve the convergence of the expansion we also
try the effect of adding the following term $\Delta H$ to $H_0$ 
and subtracting it from the perturbation $V$,
\be
\Delta H =  r' \sum_i (-1)^i S^z_{\rm tot}(i) [1- | S^z_{\rm tot} (i) | ]
\ee
where $i$ runs over rungs, and $S^z_{\rm tot}(i) $ is the total spin in
$z$-direction for rung $i$. We adjust the coefficient $r'$ to obtain best convergence.
The series have been computed to order $x^{13}$ for
the ground state energy.

In Figure 6 we present results for the ground state energy as a function of
$J/J_{\perp}$ (taking, as usual, $t_{\perp}=t$). Also shown is the curve for
the half-filled ground state energy (divided by a factor of 2).
The crossing points provides an estimate of the locus of the phase separation
line, where it becomes energetically favourable for the electrons
to cluster together. These results are plotted and discussed in the following
section. 

An interesting limiting case is where $t_{\perp} \gg t, J, J_{\perp}$.
In this limit there is a %n approximate 
mapping to 
a single $S=\case 1/2$ antiferromagnetic Heisenberg chain with $J_{\rm eff}=J/2$. 
This mapping arises from replacing the rungs with effective $S=\case 1/2$
spins, and can be established by a detailed investigation of matrix elements.
The corresponding effective Hamiltonian for this mapping is
\be
H^{\rm eff} = -t_{\perp} N/2 +
J_{\rm eff} \sum_i ( {\bf S}_i \cdot {\bf S}_{i+1}  - 1/4)
\ee

In Figure 7 we show the ground-state energy per spin versus $J/t_{\perp}$
for two choices of $(t, J_{\perp})$, together with the energy per site of the 
corresponding  chain
\be
E_0/N = -t_{\perp}/2 - J \ln 2/4
\ee
The agreement is exact near $J=0$ and quite good even up to $J/t_{\perp}=1$ 
provided $t$ and $J_{\perp}$ are small.
% for $t=0$  and any small value of $J_{\perp}$.

% Similar maping can be done for

\section{Discussion and Conclusions}
We have carried out the first comprehensive series study of the $t\mbox{-}J$ ladder system,
which is applicable to real materials as well as being of theoretical
interest in its own right. We have considered the system near half-filling
(one and two holes) and for exactly quarter filling. Our results provide a 
comparison for other numerical and analytical studies of the $t\mbox{-}J$ ladder,
especially exact diagonalizations for small systems which have
been extensively used.

Both one-hole and two-hole states propagate as quasi-particles with well defined 
excitation energies, and our estimates of these are, we believe, the
most accurate available. They are in qualitative agreement with exact diagonalization
results but the latter are for small systems and may have substantial finite size correction
terms. We have obtained precise numerical estimates for the two-hole binding energy.

a) {\bf Half-filling}

Figure 8 shows the resulting ``phase diagram" for the half-filled case.
From Figure 5 we can estimate the critical line at which the 2-hole binding
energy vanishes, shown as line (A) in Figure 8. Below the line two holes 
bind, above the line they are unbound. Also shown are the lines where the 1-hole gap 
vanishes (B), and the 2-hole gap vanishes (C). 

The binding of two holes is a necessary requirement for phase separation, where holes
clump together to form separate hole-rich and hole-poor regions; but it is
not sufficient. The usual criterion for the phase separation boundary is 
taken to be\cite{tsu94,hay96} the point where the inverse compressibility
vanishes:
\be
{1\over  n^2 \kappa } = {1\over 2 L} [ E(N+2) - 2 E(N) + E(N-2)] \label{eq_ps}
\ee
where $N$ is the number of electrons (or holes). In other words,
phase separation occurs when the energy surface becomes locally {\bf convex}
with respect to the addition or subtraction of hole pairs; whereas the binding energy
(\ref{eq_binding}) measures the convexity with respect to addition of
single holes. We are unable to compute the quantity (\ref{eq_ps}). 
Tsunetsugu {\it et al.}\cite{tsu94} and Hayward {\it et al.}\cite{hay96} use
criterion (\ref{eq_ps}) to find the phase separation boundary at line
(D) in Figure 8, very much lower than the 2-hole binding boundary (A).

This model is also believed\cite{poi95,tsu94} to exhibit a ``C1S0"
phase above the phase separation boundary with gapped spin excitations but
gapless charge excitations. Troyer {\it et al.}\cite{tsu94}
 discuss how the bound hole
pairs form a gapless  band of charge fluctuations, and how the 
dispersion relation for the pair becomes
linear near ${\bf k}=0$. The most naive criterion for the boundary
of the C1S0 phase would be the vanishing of the two-hole
gap, line (C) in Figure 8. That would imply the possibility of
a narrow intermediate region between lines (C) and (D),
corresponding to a ``C0S0" phase where phase separation has
not yet occurred. 
Other authors\cite{nis99} define the
``charge gap"
\be
\Delta_c = {1\over 2} [ E_0 (N_{\uparrow}+1, N_{\downarrow}) - 
2  E_0 (N_{\uparrow}, N_{\downarrow}) + E_0 (N_{\uparrow}-1, N_{\downarrow})]
\ee
which cannot be computed with our current techniques of series expansions.
%
% is akin to our binding energy $E_b$, equation (\ref{eq_binding}).
% This would put the boundary of the C1S0 region right up at line (A),
% inconsistent with the discussion of Troyer {\it et al.}\cite{tsu94}.

b) {\bf Quarter-filling}

From Figure 6, we can estimate the boundary at which
\be
E_0 (n=\case 1/2) = E_0(n=1)/2
\ee
where $n$ is the number of electrons per site ($n=1$ for half-filling, $n=1/2$
for quarter-filling).
This is a sufficient but not necessary criterion for phase separation,
implying a `global" convexity of the energy surface, so that the system
could save energy by separating completely into a hole-empty and
hole-full region.
Estimates based on this criterion are given by the dotted line in Fig. 9.
 Estimates based on the inverse 
compressibility\cite{tsu94} are shown by the open points in Figure 9:
they lie a little above our estimates, but quite close to it,
and are hardly distinguishable within present accuracy limits.

We have noted on a mapping (exact at the limit $t,J,J_{\perp} \ll t_{\perp}$)
from the quarter-filled  $t\mbox{-}J$ ladder to an $S=\case 1/2$ antiferromagnetic Heisenberg spin chain.
This idea can be  generalized to other systems, such as the dimerized $t\mbox{-}J$ chain or
dimerized $t\mbox{-}J$ model on a square lattice, and we will report on this in future work.

% Precise numerical estimates have been obtained for the two-hole
% binding energy. From these results (Fig. 5) we estimate the critical
% line which separates the binding and non binding regions in the plane of
% $t/J_{\perp}$ versus $J/J_{\perp}$. This is shown in Fig. 8. In the
% same figure we show the location of the phase separation line
% for quarter filling, as discussed in the previous section.

Note added in proofs: After submission we learnt of recent work of 
Bose and Gayen\cite{bos99} in which they obtain exact results for
1 and 2-hole states in a $t-J$ ladder with additional diagonal
interactions. The physics of two systems appears similar.

\acknowledgments
We are grateful to O. Sushkov and M. Troyer for  useful discussions.
This work forms part of a research project supported by a grant
from the Australian Research Council. The computation has been performed
on Silicon Graphics Power Challenge and Convex machines. We thank the New
South Wales Centre for Parallel Computing for facilities and assistance
with the calculations. 

\appendix
\section{}
If we take the term in Eq. (\ref{h07}) as the unperturbed Hamiltonian, and
all the remaining terms as the perturbation, we can get a 
double series for the excitation energies. The series 
up to order $t^i J^j$ ($i+j \leq 4$) for 
the excitation spectra of the one-hole symmetric state $\Delta E^{1S}$ is 
(with $J_{\perp}=1$  and $t_{\perp}=t$)
\bea
\Delta E^{1S} &=& 1 + (J/2 - t) + (3J^2 /8 -3 t^2/2) + 
( 3 J^3/16 + 3 t J^2/4 - 9 t^2 J/4 + 3 t^3 ) \nonumber \\
&& + (-3 J^4 /32 + 3 t J^3/4 - 69 t^2 J^2 /32 + 9 t^3 J -105 t^4/16 ) \nonumber \\
&& + (  t - 3 t J/2 - 9 t J^2/8 + 3 t^2 J - 9 t^3/4 + 9 t J^3 /32 
+ 9 t^2 J^2/2 - 15 t^3 J/2 + 9 t^4 ) \cos( k_x)  \nonumber \\
&& + ( 15t^2 J^2/16  + 3 t^3 J/2 - 3 t^4/2 ) \cos(2 k_x) 
+ \cdots
\eea

The series for the one-hole antisymmetric state is related to that for the
symmetric case by
\bea
\Delta E^{1A} (t,J,k_x) = \Delta E^{1S} (- t,J,\pi-k_x)
\eea

The two-hole excitation energy  $\Delta E^{2h}$ is 
\bea
\Delta E^{2h} &=& 
1 + J + 3 J^2/4 - 4 t^2 + 3 J^3/8 -3 J^4/32 + 3 t^2 J^2/8 + t^4/2 \nonumber\\
&& + ( -4 t^2 +3 t^2 J^2/4 + 6 t^4 ) \cos(k_x)
+ 11 t^4 \cos(2 k_x)/2
+\cdots
\eea

% \newpage

\begin{figure} %1
\caption{Stucture and parameters of the $t\mbox{-}J$ ladder. 
}
\label{fig_1}
\end{figure}

\begin{figure} %2
\caption{Excitation spectra for one-hole symmetric and antisymmetric states for
$J/J_{\perp}=0,0.5,1.0$ and various $t/J_{\perp}$. In each case the 
dispersionless energy for $t=0$ is also shown.
}
\label{fig_2}
\end{figure}

\begin{figure} %3
\caption{Comparison of our results (points connected by dashed line)
for the one-hole symmetric spectrum with exact 
diagonalizations (Troyer {\it et al.}\protect\cite{tsu94}, solid curve) and an approximate analytic
theory (Sushkov\protect\cite{sus98}, dotted curves).
}
\label{fig_3}
\end{figure}

\begin{figure} %4
\caption{Excitation spectra for two-hole  states for
$J/J_{\perp}=0,0.5,1.0$ and various $t/J_{\perp}$. In each case the 
dispersionless energy for $t=0$ is also shown.
}
\label{fig_4}
\end{figure}

\begin{figure} %5
\caption{Binding energy of two holes vs $t/J_{\perp}$.
The $J=0$ case shows estimates for different integrated differential
approximants, which begin to spread out for $t/J_{\perp}\protect\gtrsim 1$.
The points where $E_b=0$ give the two-hole binding line shown in 
Fig. \protect\ref{fig_8}. Also shown are the results of Hayward {\it et al.}\protect\cite{hay962}
for the isotropic case  $J/J_{\perp}=1$ (the big full points connected by
the dot-long dash line).
}
\label{fig_5}
\end{figure}

\begin{figure} %6
\caption{Ground state energy per site for quarter-filling versus $J/J_{\perp}$
for various $t/J_{\perp}$. The solid line is the ground state energy
for the phase separated state. The crossing points yield the phase separation line
shown in Fig. \ref{fig_9}.
}
\label{fig_6}
\end{figure}

\begin{figure} %7
\caption{Comparison of the ground state energy per site for quarter-filling
for the case of $t=0$ (note that the results are independent of $J_{\perp}$ for 
a small value of $J_{\perp}$) and the case of $J_{\perp}=t=J$
with the results of an approximate mapping to a Heisenberg spin chain (solid line).
}
\label{fig_7}
\end{figure}

\begin{figure} %8
\caption{Phases and critical lines for the $t\mbox{-}J$ ladder at half-filling, 
in the plane of $t/J_{\perp}$ versus $J/J_{\perp}$ (see text). Two holes do not bind in the 
region marked NB.
}
\label{fig_8}
\end{figure}

\begin{figure} %9
\caption{Phase separation boundary at quarter-filling in the $t/J_{\perp}$
versus $J/J_{\perp}$ plane. The region marked PS is where phase separation occurs. 
The open points mark the boundary value found by Hayward and Poiblanc\protect\cite{tsu94} for
periodic (PBC) and open (OPC) boundary conditions on 8-rung finite chains.
}
\label{fig_9}
\end{figure}

%\setdec 0.0000000000
\begin{table}
%\squeezetable
\caption{The nine rung states and their energies.}\label{tab1}
\begin{tabular}{|c|c|c|c|}
\multicolumn{1}{|l|}{No.} &\multicolumn{1}{l|}{Eigenstate}
&\multicolumn{1}{l|}{Eigenvalue} &\multicolumn{1}{l|}{Name} \\
\hline
1  &  $\frac{1}{\sqrt 2}(\mid \uparrow \downarrow \rangle - \mid \downarrow \uparrow \rangle )$ &    $-J_{\perp}$  &  singlet  \\
\hline
2  & $\mid \downarrow \downarrow \rangle  $  &    $0$  &  triplet ($S^z_{\rm tot}=-1$)  \\
\hline
3  & $\frac{1}{\sqrt 2}(\mid \uparrow \downarrow \rangle + \mid \downarrow \uparrow \rangle )$ & $0$  &  triplet ($S^z_{\rm tot}=0$)  \\
\hline
4  & $\mid \uparrow \uparrow \rangle  $  &    $0$  &  triplet ($S^z_{\rm tot}=1$)  \\
\hline
5  & $\mid 00 \rangle$  &    $0$  &  hole-pair singlet  \\
\hline
6  & $\frac{1}{\sqrt 2}(\mid 0 \downarrow \rangle + \mid \downarrow 0 \rangle )$ & $-t_{\perp}$  & electron-hole bonding ($S^z_{\rm tot}=-\case 1/2$)  \\
\hline
7  & $\frac{1}{\sqrt 2}(\mid 0 \uparrow \rangle + \mid \uparrow 0 \rangle )$ & $-t_{\perp}$  & electron-hole bonding ($S^z_{\rm tot}=\case 1/2$)  \\
\hline
8  & $\frac{1}{\sqrt 2}(\mid 0 \downarrow \rangle - \mid \downarrow 0 \rangle )$ & $t_{\perp}$  & electron-hole antibonding ($S^z_{\rm tot}=-\case 1/2$)  \\
\hline
9  & $\frac{1}{\sqrt 2}(\mid 0 \uparrow \rangle - \mid \uparrow 0 \rangle )$ & $t_{\perp}$  & electron-hole antibonding ($S^z_{\rm tot}=\case 1/2$)  \\
\end{tabular}
\end{table}


\begin{references}
\bibitem[*]{byline1}e-mail address: j.oitmaa@unsw.edu.au
\bibitem[\dag]{byline2}e-mail address: c.hamer@unsw.edu.au
\bibitem[\ddag]{byline3}e-mail address: w.zheng@unsw.edu.au

%1
\bibitem{azu94}M. Azuma, Z. Hiroi, M. Takano, K. Ishida and
Y. Kitaoka, Phys. Rev. Lett. {\bf 73}, 3463(1994).

\bibitem{hir95}Z. Hiroi and M. Takano, Nature {\bf 377}, 41(1995).

%3
\bibitem{dag96}E. Dagotto and T.M. Rice, Science  {\bf 271}, 618(1996).

\bibitem{ric97}T.M. Rice, Z. Phys. B{\bf 103}, 165(1997).

%5
\bibitem{mar96}M.A. Martin-Delgado, R. Shankar and G. Sierra, Phys.
Rev. Lett. {\bf 77}, 3443(1996).

\bibitem{ladder}W.H. Zheng, V. Kotov, and J. Oitmaa, Phys. Rev. B {\bf 57}, 
11439(1998).

%7
\bibitem{ner98}A.A. Nersesyan, A.O. Gogolin and F.H.L. Essler, Phys. Rev. Lett.
 {\bf 81}, 910(1998).

\bibitem{cab}D.C. Cabra and M.D. Grynberg, Phys. Rev. Lett. {\bf 82}, 1768(1999).

%9
\bibitem{kot99}V.N. Kotov, O.P. Sushkov and R. Eder, Phys. Rev. B {\bf 59}, 6266(1999).

%10
\bibitem{dag92}E. Dagotto, J. Riera and D.J. Scalapino, Phys. Rev. B {\bf 45}, 
5744(1992).

\bibitem{tsu94}H. Tsunetsugu, M. Troyer and T.M. Rice, Phys. Rev. B {\bf 49},
16078(1994); M. Troyer, H. Tsunetsugu and T.M. Rice, {\it ibid.} B {\bf 53}, 251(1996).

%12
\bibitem{poi95}D. Poilblanc, D.J. Scalapino and W. Hanke, Phys. Rev.
B {\bf 52}, 6796(1995).

\bibitem{hay96}C.A. Hayward and D. Poilblanc, Phys. Rev. B {\bf 53}, 11721(1996).

\bibitem{hay962}C.A. Hayward,  D. Poilblanc and D.J. Scalapino, Phys. Rev. B {\bf 53}, R8863(1996).

%14
\bibitem{mul98}T.F.A. M\"uller and T.M. Rice, Phys. Rev. B {\bf 58}, 3425(1998).

%15
\bibitem{rie99}J. Riera, D. Poilblanc and E. Dagotto, Eur. Phys. J B {\bf 7}, 53(1999).

%16
\bibitem{whi97}S.R. White and D.J. Scalapino, Phys. Rev. B {\bf 55}, 6504(1997).

%17
\bibitem{sie98}G. Sierra, M.A. Martin-Delgado, J. Dukelsky, S.R. White and
D.J. Scalapino, Phys. Rev. B {\bf 57}, 11666(1998).

%18
\bibitem{sig94}M. Sigrist, T.M. Rice and F.C. Zhang, Phys. Rev. B {\bf 49}, 
12058(1994).

%19
\bibitem{lee}Y.L. Lee, Y.W. Lee, C.Y. Mou and Z.Y. Weng, preprint cond-mat/9903059.

%20
\bibitem{sus98}O.P. Sushkov, Phys. Rev. B {\bf 60}, 3289(1999).

\bibitem{bal96}L. Balents and M.P.A. Fisher, Phys. Rev. B {\bf 53}, 12133 (1996).

%19
\bibitem{ham98}C.J. Hamer, W.H. Zheng and J. Oitmaa, Phys. Rev. B {\bf 50}, 15508(1998).

\bibitem{bar92}T. Barnes, A.E. Jacobs, M.D. Kovarik, and W.G. Macready, Phys. Rev.
  B {\bf 45}, 256(1992), and references therein.

%20
\bibitem{gut}A.J. Guttmann, in {\it Phase Transitions and Critical Phenomena},
edited by C. Domb and M.S. Green (Academic, New York, 1989), Vol. 13.

%\bibitem{gel90}
%M. P. Gelfand, R. R. P. Singh and D. A. Huse,
%J. Stat. Phys. {\bf 59}, 1093 (1990).

\bibitem{gelfand}M. P. Gelfand,
Solid State Comm. {\bf 98}, 11 (1996).

\bibitem{smo98}H. Smolinski {\it et al.} Phys. Rev. Lett. {\bf 80}, 5164(1998).

\bibitem{nis98}S. Nishimoto and Y. Ohta, J. Phys. Soc. Japan {\bf 67}, 2996(1998);
   {\it ibid} {\bf 67}, 4010(1998).

\bibitem{nis99}S. Nishimoto and Y. Ohta, Phys. Rev. B {\bf 59}, 4738(1999).

%\bibitem{whi94}White, Noack and Scalapino, Phys. Rev. Lett. {\bf 73}, 886(1994).

\bibitem{bos99}I. Bose and S. Gayen,  J. Phys. Condens. Matter {\bf 11}, 6427(1999).
\end{references}
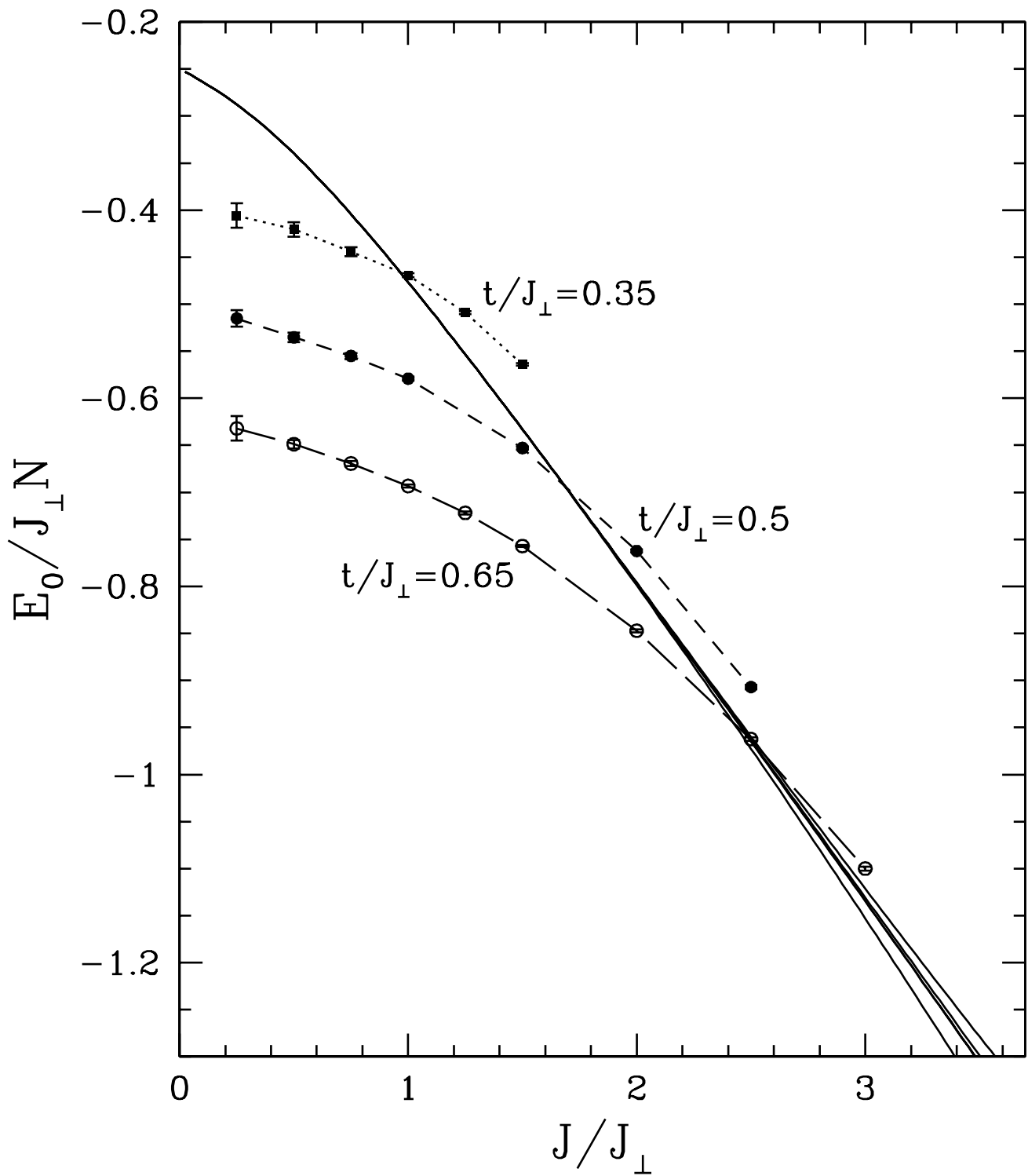
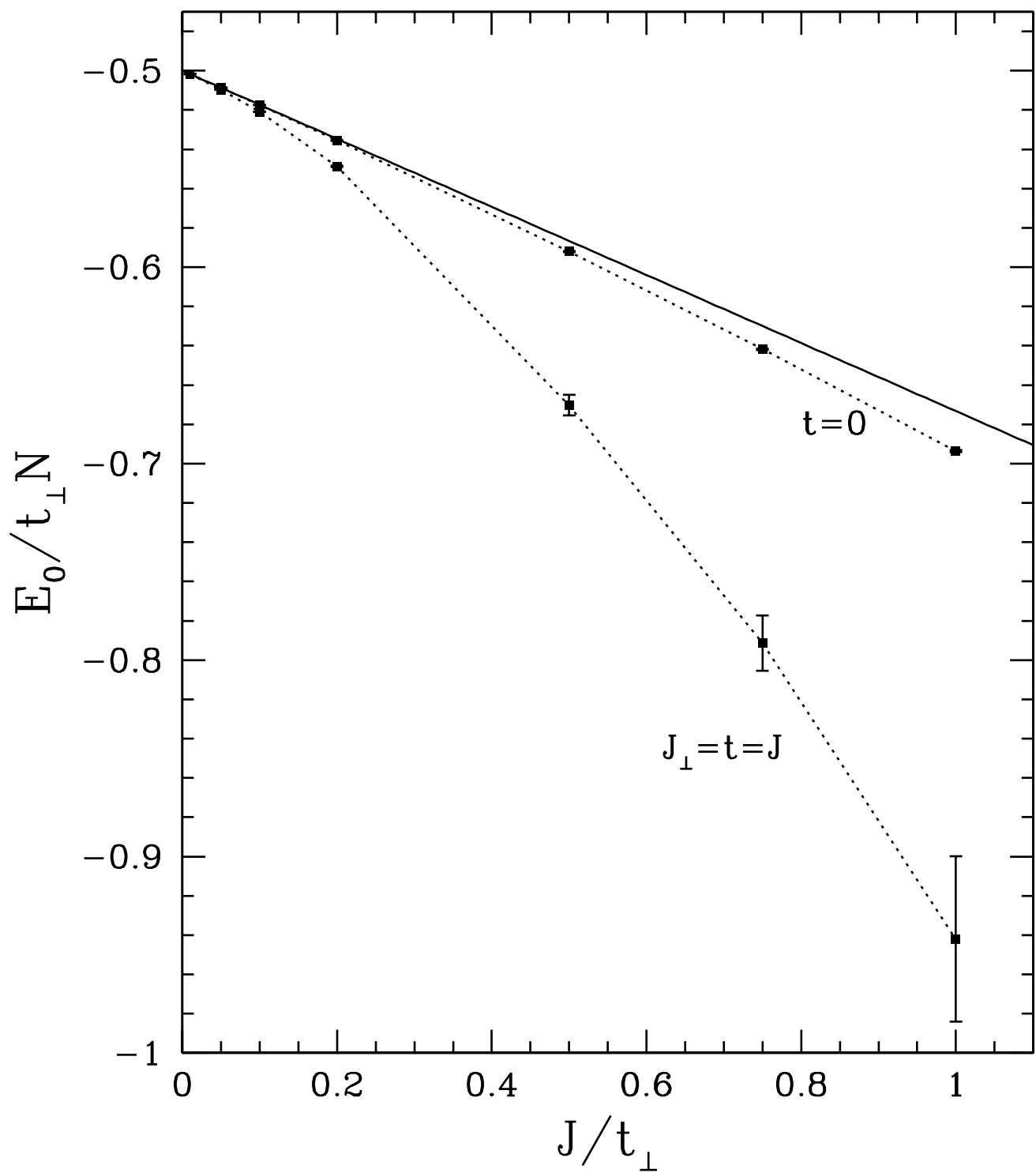
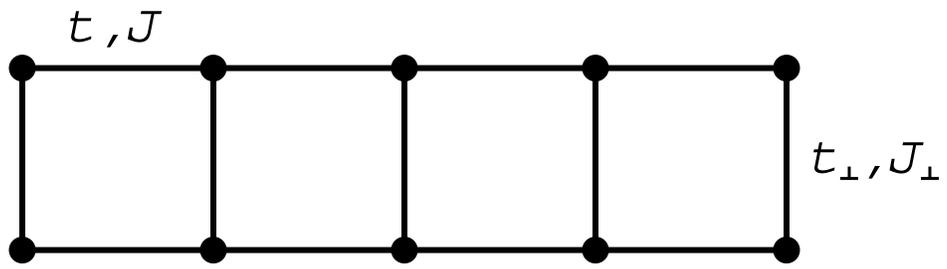
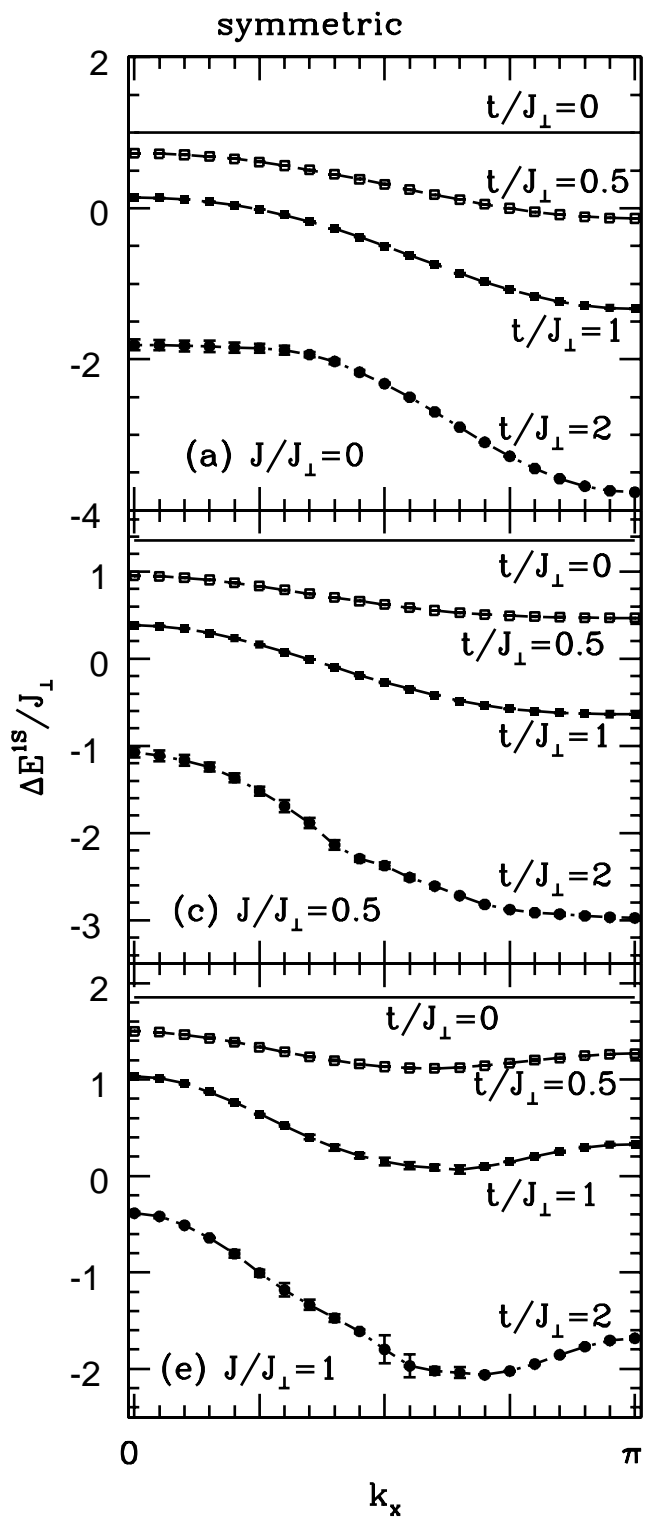
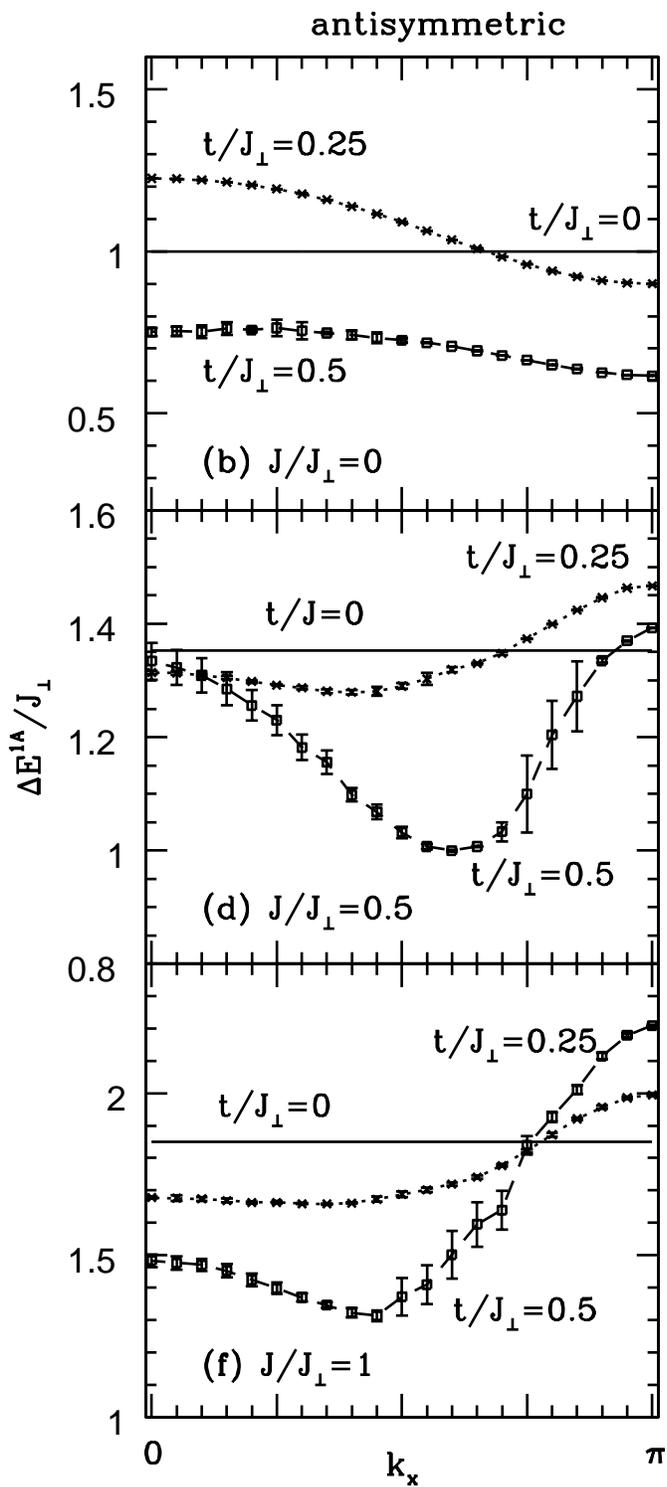
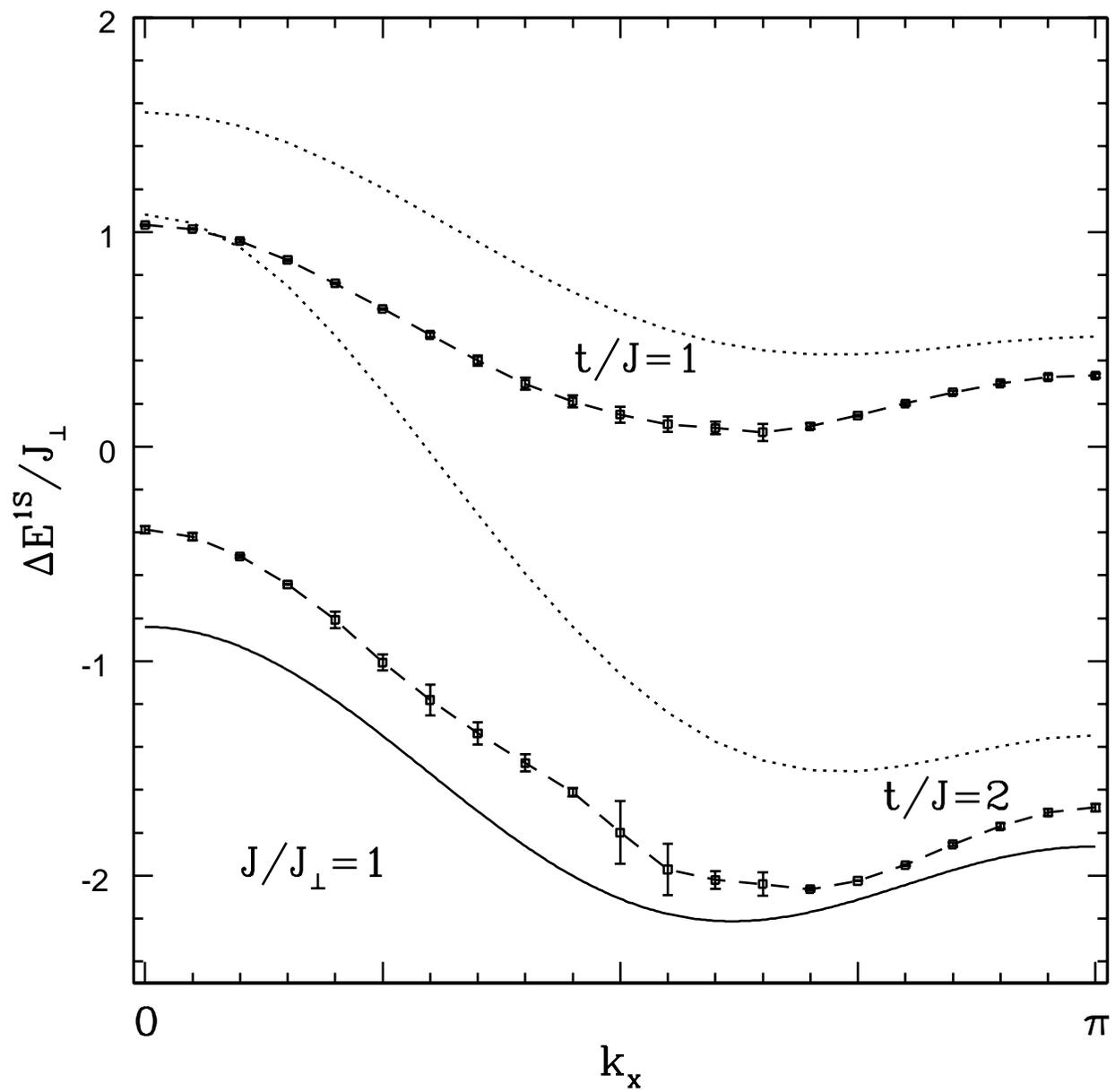
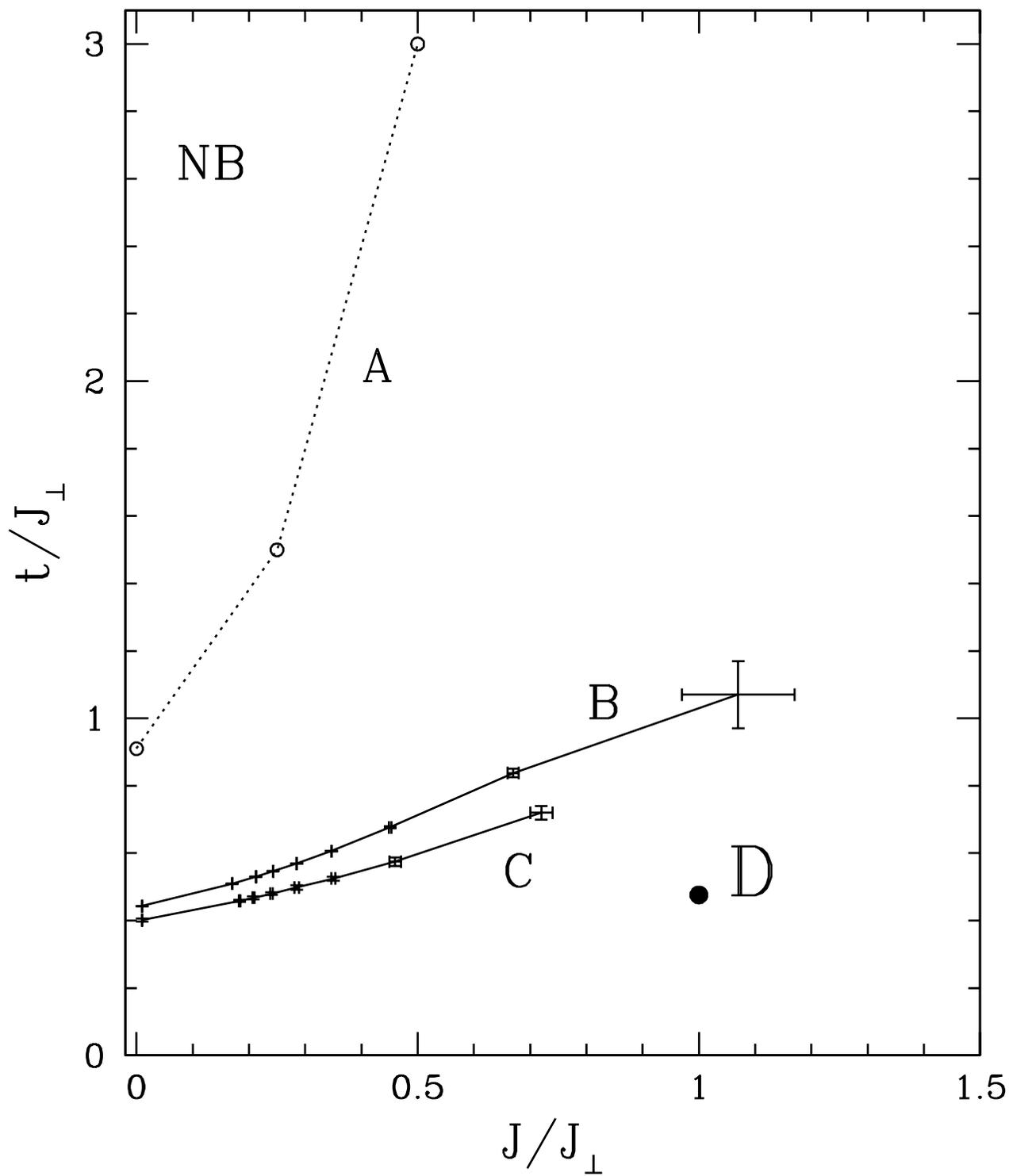
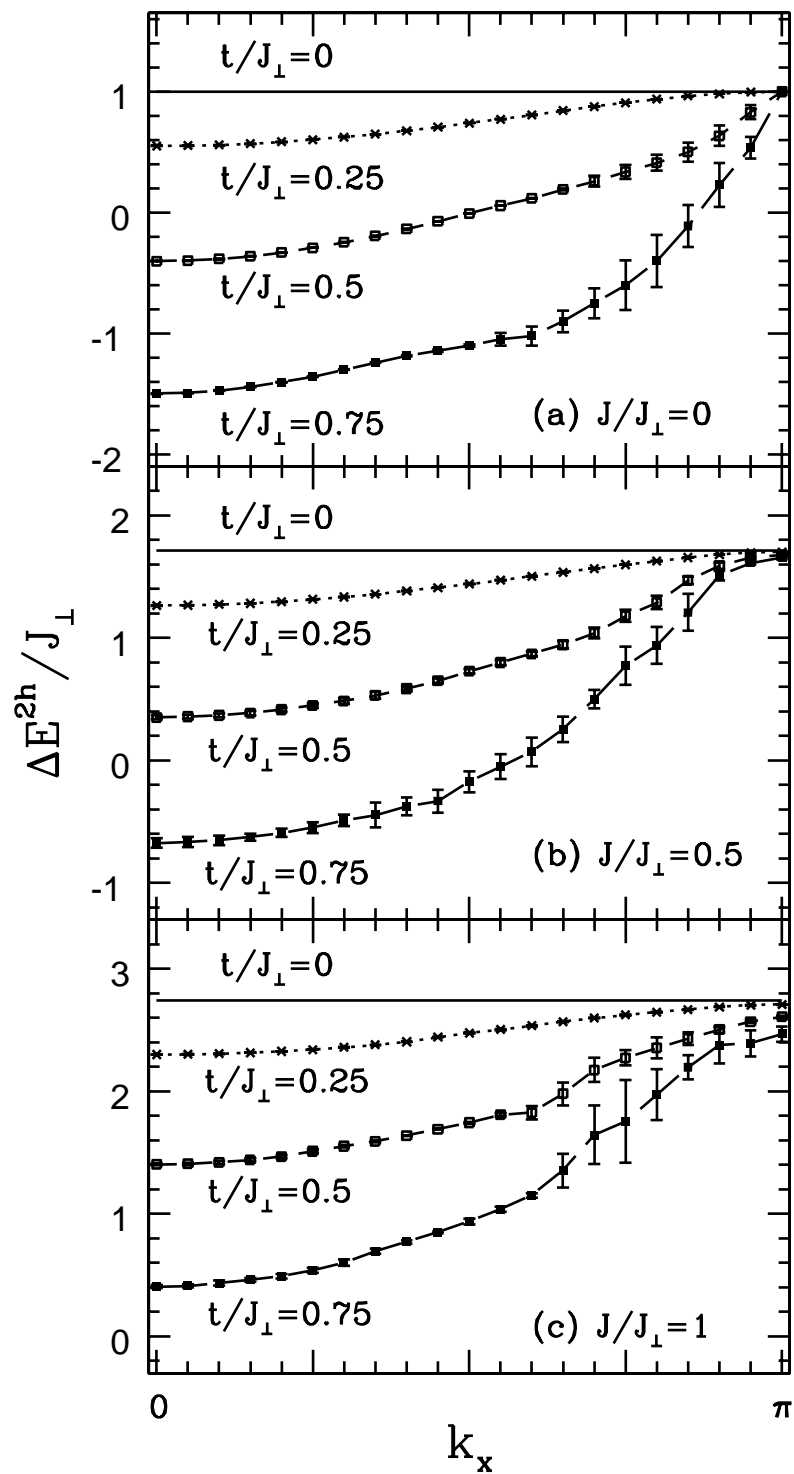
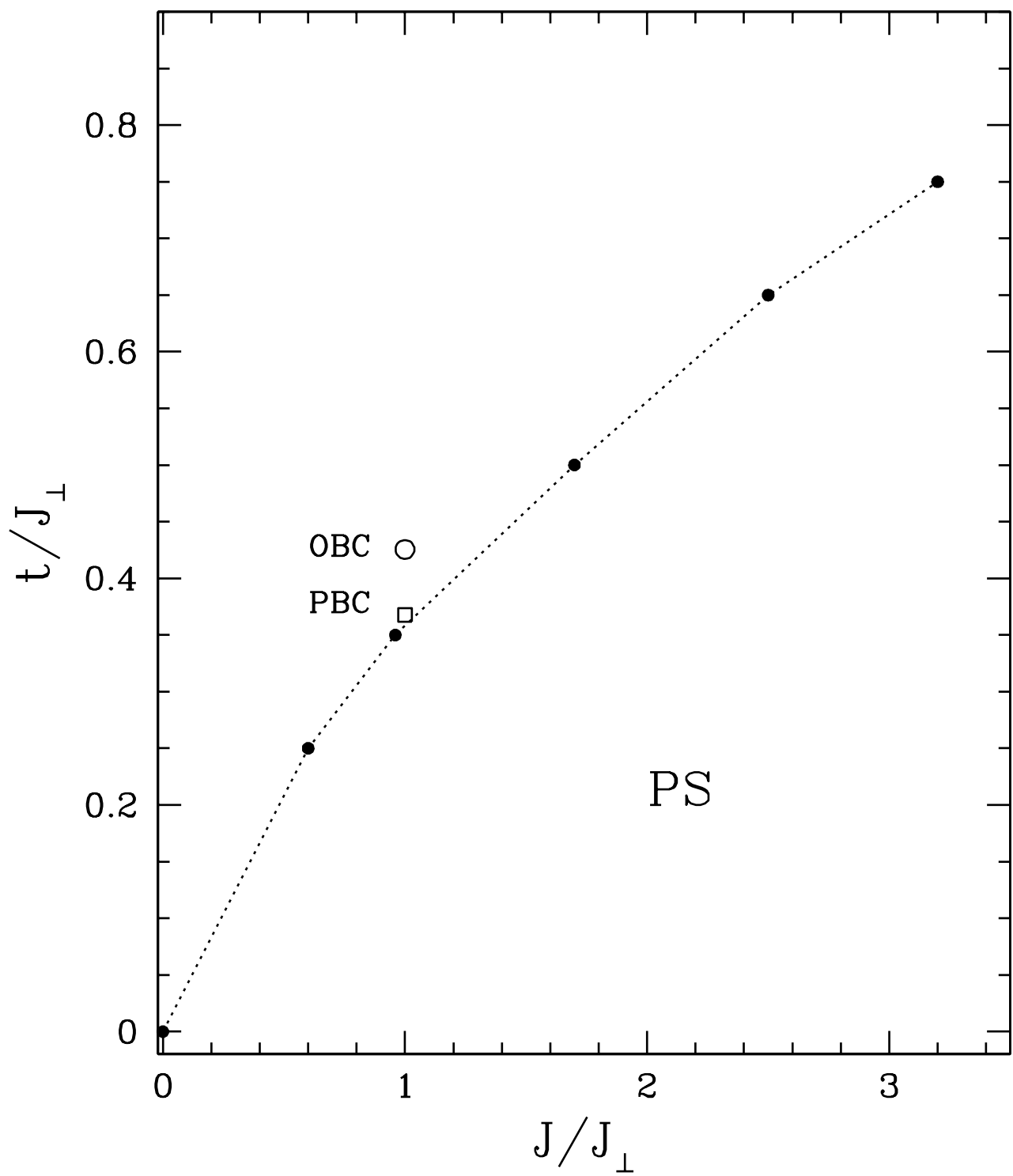
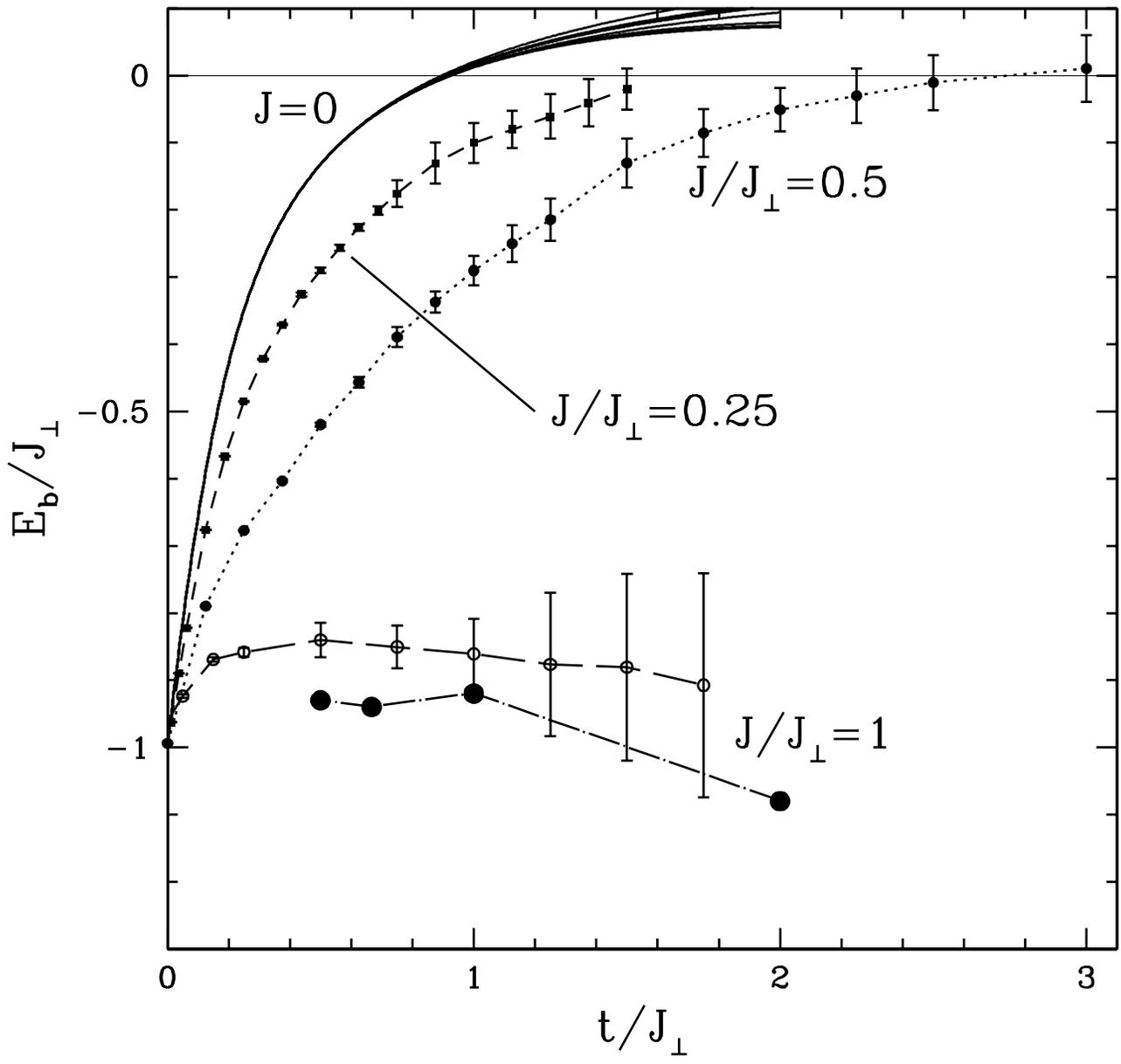
\end{document}